\documentclass[pre,twocolumn,byrevtex,superscriptaddress,pdftex]{revtex4}
\usepackage{amsmath,epsfig,enumerate,graphicx}
\usepackage{verbatim}
\usepackage{amstext}
\usepackage{amsthm}
\usepackage{ifthen}
\newboolean{twocolswitch}
\newboolean{revtexswitch}

\def\d{{\rm d}}

\setboolean{twocolswitch}{true}
\setboolean{revtexswitch}{true}

\begin{document}

\title{
Post-death Transmission of Ebola:\\
Challenges for Inference
and Opportunities for Control
}

\author{
  \firstname{Joshua S.}
  \surname{Weitz}
  }
\email{jsweitz@gatech.edu}
\homepage{http://ecotheory.biology.gatech.edu}
\affiliation{
   	School of Biology,
	Georgia Institute of Technology,
	Atlanta, GA, USA
	}
\affiliation{
   	School of Physics,
	Georgia Institute of Technology,
	Atlanta, GA, USA
}
\author{
  \firstname{Jonathan}
  \surname{Dushoff}
  }
\affiliation{
   	Department of Biology,
	McMaster University,
	Hamilton, ON, Canada
	}
\affiliation{
   	Institute of Infectious Disease Research,
	McMaster University,
	Hamilton, ON, Canada
}

\date{\today}

\begin{abstract}
Multiple epidemiological models have been proposed to predict the spread of Ebola in West Africa.
These models include consideration of counter-measures meant to slow and, eventually, stop the spread of the disease.
Here, we examine one component of Ebola dynamics that is of growing concern -- the transmission of Ebola from the dead to the living.  
We do so by applying the toolkit of mathematical epidemiology to analyze the consequences of post-death transmission.
We show that underlying disease parameters cannot be inferred with confidence from early-stage incidence data (that is, they are not ``identifiable'') because different parameter combinations can produce virtually the same epidemic trajectory. 
Despite this identifiability problem, we find robustly that inferences that don't account for post-death transmission tend to underestimate the basic reproductive number -- thus, given the observed rate of epidemic growth, larger amounts of post-death transmission imply larger reproductive numbers. 
From a control perspective, we explain how
improvements in reducing post-death
transmission of Ebola may
reduce the overall epidemic spread and scope substantially.
Increased attention to the proportion of post-death transmission has the potential to aid both in projecting the course of the epidemic and in evaluating
a portfolio of control strategies.
\end{abstract}

\maketitle

\section*{Introduction}
A recent, influential modeling paper concluded, 
based on data available as of September 2014,
that the ongoing Ebola epidemic
in Guinea, Liberia and Sierra Leone had the potential to exceed $1$ million new cases by mid-January 2015, in the absence of intervention~\cite{meltzer_2014}.
Even with intervention and changes in behavior,
a follow-up study by an independent group in October 2014 estimated
that $100,000$ additional cases could be expected in Liberia alone by
mid-December 2014, unless a coordinated,
large-scale response is implemented rapidly~\cite{lewnard_2014}.
These predictions leveraged the structure of previous
epidemiological models~\cite{chowell_2004,legrand2007understanding}
that encapsulate the 
infection cycle of Ebola virus disease (EVD), by tracking
the dynamics and interactions of different types of individuals within
a population including Susceptible, Exposed, Infectious and Removed types. 
Exposed individuals
are infected but not yet infectious (i.e., also referred to as
latently infected).
In a SEIR model representation of EVD dynamics,
the R class accounts for two types of individuals: those who
recovered from the disease and those who have died from the disease
(and are therefore ``removed'').

However, a complication in modeling EVD arises because
EVD may be contracted by direct contact with bodily fluids from
individuals who are alive and from those who have died from the 
disease~\cite{francesconi_2003,CDCEbolaQA2014}.
In the present epidemic, contact tracing of 701 individuals confirmed to have been infected with EVD in the ongoing epidemic found that 67 patients reported contacts with individuals who died of EVD, but not with any living EVD cases, while 148 patients reported contacts with both living and dead individuals infected with EVD~\cite{teamebola}, consistent with $10\%=\frac{67}{701}$ to $30\%=\frac{67+148}{701}$ of
Ebola cases being caused by post-death transmission.  
If funerals and burial rites can act as ``super-spreader''
events~\cite{lloyd2005,galvani2005,pandey_2014}, the true fraction may lie outside this range:
for example, Legrand and colleagues~\cite{legrand2007understanding} estimated
that 2/3 of the total ${\cal{R}}_0$ for the 1995 EVD outbreak
in the Democratic Republic of Congo could be attributed to post-death
transmission.  
Some early~\cite{legrand2007understanding}
and recent (e.g.,~\cite{lewnard_2014,pandey_2014,gomes2014,rivers_2014})
epidemiological models of EVD
have incorporated a D class, thereby distinguishing between
recovered and dead individuals.  Other models treat post-death
transmission implicitly, by increasing the effective transmission rate
and/or duration in the I class~\cite{meltzer_2014,althaus2014}.  
Yet, the implications of post-death
transmission for inferences about epidemic spread have not been evaluated systematically.  As we show,
uncertainty in the relative force of infection before- and after-death
has a number of consequences for estimating ${\cal{R}}_0$
and the potential for control of the ongoing Ebola epidemic.

\begin{figure*}[bt!]
\begin{center}
\includegraphics[width=0.75\textwidth]{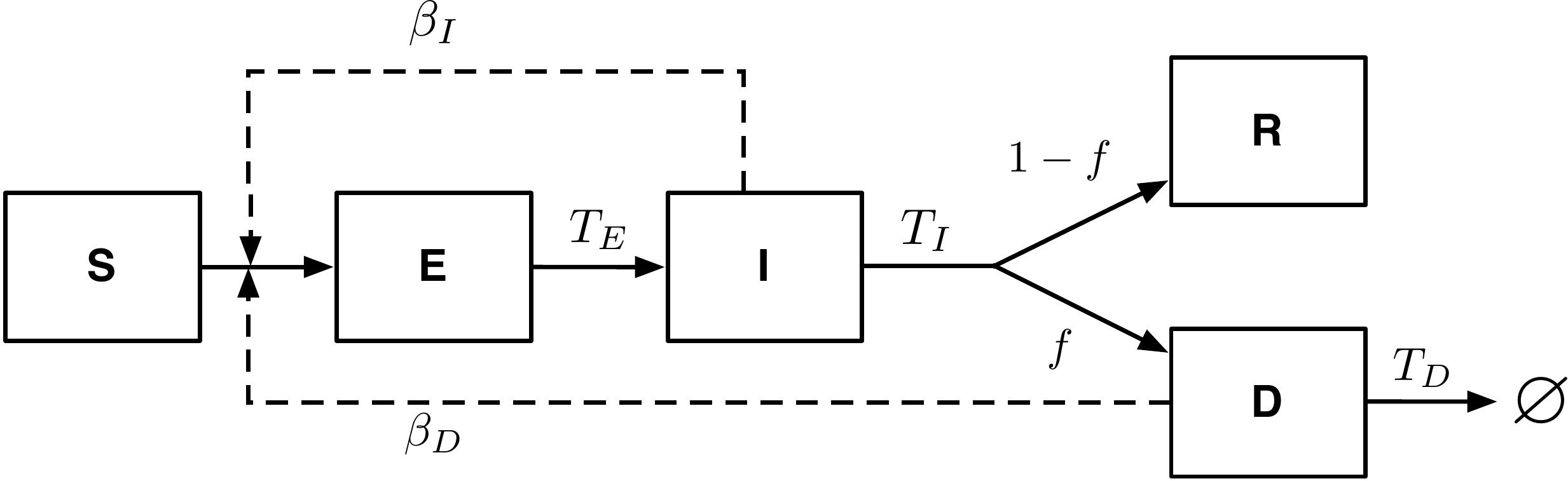}
\caption{Schematic of the SEIRD model, i.e., the dynamics of
Susceptible, Exposed (i.e., latently infected), Infectious, Recovered and Dead (but still infectious)
individuals.  Solid arrows denote transitions between states.  Dashed 
arrows denote that transmission depends on interactions between $S$ and
$I$ individuals or between $S$ and $D$ individuals.  
Parameters $\beta_I$ and $\beta_D$ are transmission rates, $T_E$, $T_I$
and $T_D$ are the average periods in the $E$, $I$ and $D$ class,
respectively, and $f$ is the
fraction of individuals who die of EVD.
\label{fig.seird}
}
\end{center}
\end{figure*}

\section*{Results}
\subsection*{The basic reproductive number, ${\cal{R}}_0$, of EVD
includes the effects of post-death transmission}
The basic reproductive number, ${\cal{R}}_0$, denotes the
average number of secondary cases caused by a single
infected individual in an otherwise susceptible population.
The criterion for epidemic spread in standard epidemiological
models is that ${\cal{R}}_0>1$ so that the initial infection gives rise,
on average, to more than one infected case.  

In a conventional SEIR model, EVD transmission between
infected and susceptible individuals occurs at an average rate $\beta_I$ over a period of infectiousness $T_I$. 
A fraction $f$ of infected individuals die and the remainder, $1-f$,
recover and are assumed to be permanently immune to subsequent infection.
A SEIRD model includes an additional transmission route:
dead individuals can transmit EVD to susceptible individuals
at a rate $\beta_D$ 
over a period of infectiousness $T_D$, after which they are
permanently removed from the system via burial or loss of infectiousness
(see Figure ~\ref{fig.seird}).  
Appendix~\ref{app.seird_model} provides the mathematical details
of the model.
The basic reproductive number in the SEIRD model is:
\begin{equation}
{\cal{R}}_0 = \beta_I T_I+f\beta_DT_D
\end{equation}
The first term denotes the average number of secondary
infections due to contact with an infected individual before-death.
The second term denotes the average number of secondary
infectious due to contact with an infected individual after-death.
The number of cases arising from contact
with dead individuals is
modulated by the fraction, $f$, of infected individuals that die
due to EVD.
In contrast, the basic reproductive number in the SEIR model is:
\begin{equation}
{\cal{R}}_{0,SEIR} = \beta_{\mathrm SEIR} T_{\mathrm SEIR}
\end{equation}
It might seem that the basic reproductive number
of a SEIRD model should exceed that of a SEIR model. 
In fact, this will depend on how parameters are estimated. If the SEIR model is fit from data, then $\beta_{\mathrm SEIR}$ and $T_{\mathrm SEIR}$ will reflect transmission from both living and dead infectious individuals.
Thus, we ask: What is the change in
the estimated value of ${\cal{R}}_0$ given alternative
model frameworks meant to explain the same infected case data?

\begin{figure*}[bt!]
\begin{center}
\includegraphics[width=0.95\textwidth]{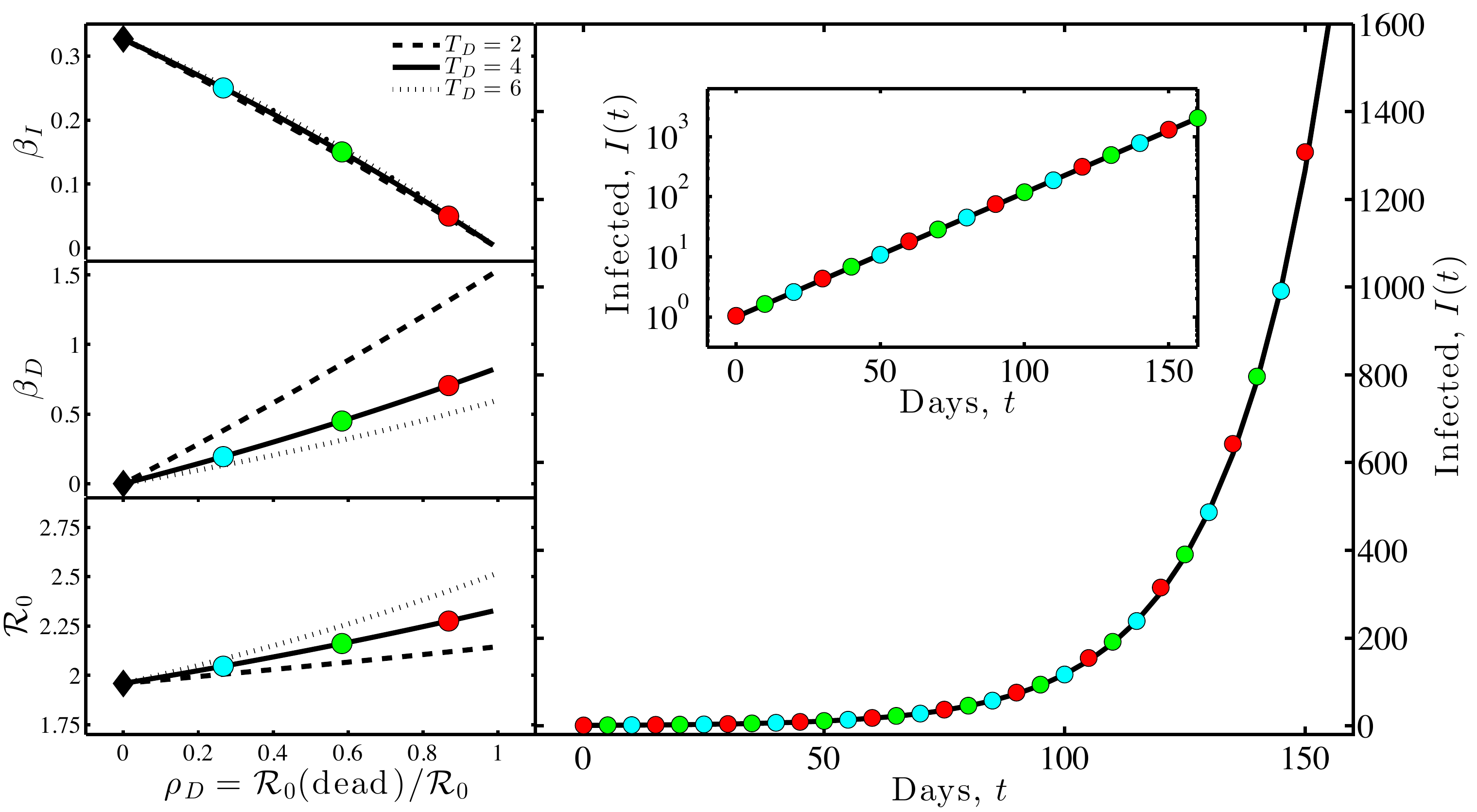}
\caption{Identifiability problems in inferring epidemiological
parameters from early-stage epidemic growth data.
(Model) The dynamics are that of an SEIR model (diamond) and
SEIRD model (lines and circles) with
epidemiological parameters $T_E=11$, $T_I=6$ and $f=0.7$.  For the SEIRD
model, three scenarios are considered where $T_D=2$, 4 and 6 days. 
For each scenario, we find various combinations of $\beta_D$ and $\beta_I$
that lead to the same early-stage epidemic
growth rate, $\lambda=1/21$. 
(Left) The calculated values of $\beta_I$, $\beta_D$ and
${\cal{R}}_0$ (from top-to-bottom) as a function of $\rho_D = {\cal{R}}_0(\mathrm{dead})/{\cal{R}}_0$.
(Right panel) Dynamics of epidemics for the three highlighted
scenarios corresponding to $\beta_I=0.05$, 0.15 and 0.25 and
$\beta_D=0.71$, 0.45 and 0.20 respectively.  
\label{fig.identify_SEIRD}
}
\end{center}
\end{figure*}
\begin{figure*}[h!]
\begin{center}
\includegraphics[width=0.67\textwidth]{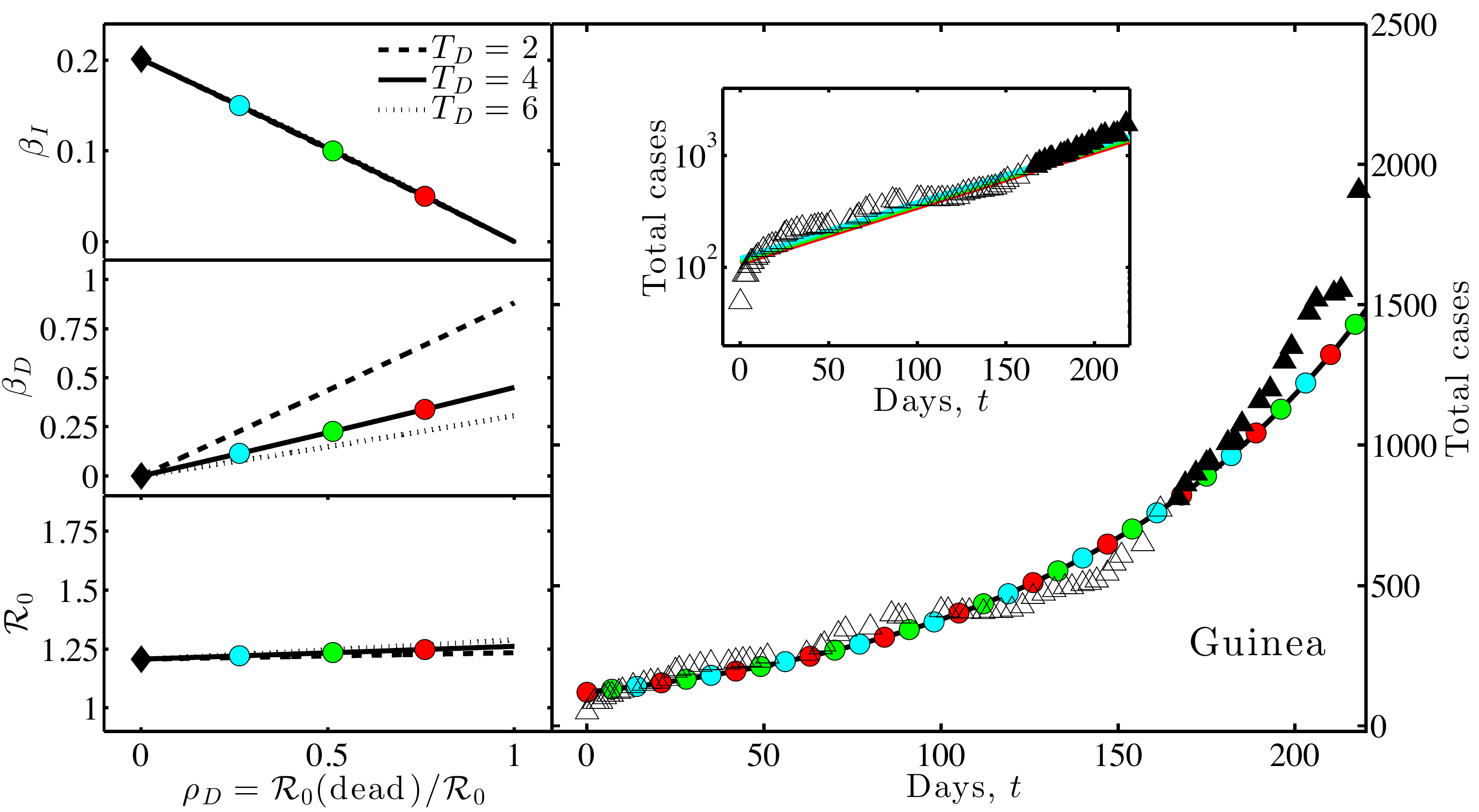}
\includegraphics[width=0.67\textwidth]{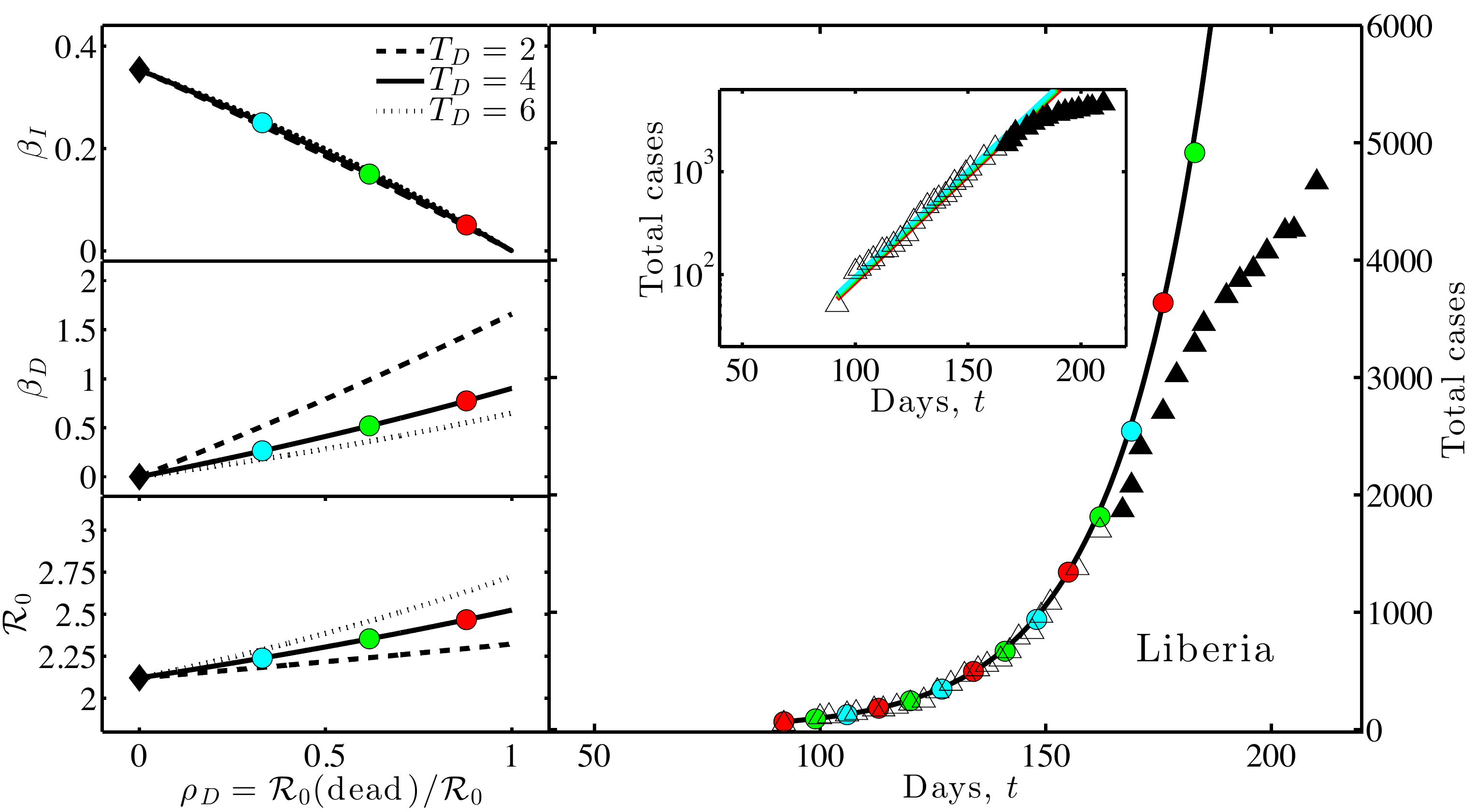}
\includegraphics[width=0.67\textwidth]{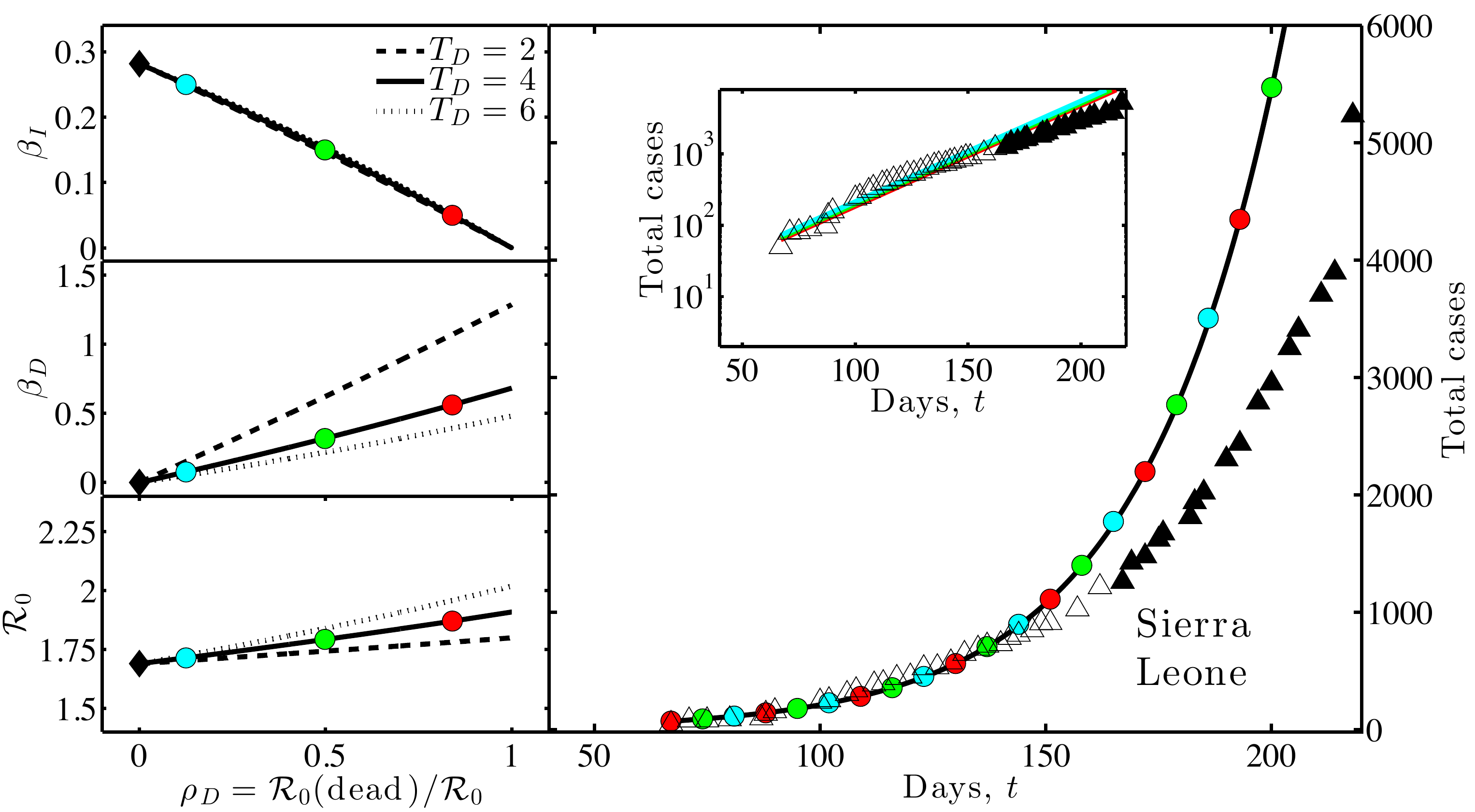}
\caption{Inferring epidemiological
parameters from early-stage Ebola epidemic case data in West Africa.
(Data) A portion of cumulative case data~\cite{rivers_github} was used
to identify an exponential growth rate $\hat{\lambda}$ (see Appendix~\ref{app.sec.data}).
The calibration regime for model fits is denoted in open triangles, where
Day 50 is 5/11/14, Day 100 is 6/30/14, Day 150 is 8/19/14 and Day 200
is 10/8/14.
(Model) The dynamics are that of an SEIR model (diamond) and
SEIRD model (lines and circles) with
epidemiological parameters $T_E=11$ (modeled as a gamma distribution), 
$T_I=6$ and $f=0.7$.  For the SEIRD
model, three scenarios are considered where $T_D=2$, 4 and 6 days. 
(Left) The calculated values of $\beta_I$, $\beta_D$ and
${\cal{R}}_0$ (from top-to-bottom) as a function of $\rho_D = {\cal{R}}_0(\mathrm{dead}~)/{\cal{R}}_0$.
(Right panel) Dynamics of epidemics for the three highlighted
scenarios.  For each country, all three scenarios lead to the same country-specific
exponential epidemic growth rate, $\hat{\lambda}$.
\label{fig.identify_SEIRD_data}
}
\end{center}
\end{figure*}

\subsection*{Identifiability problems in estimating the basic reproductive
number, ${\cal{R}}_0$}
The SEIRD model, like the SEIR, SIR and other epidemiological models,
predicts that there should be an
exponential increase in the number 
of infected cases, i.e., $I(t)\sim e^{\lambda t}$, after
an initial transient phase and before interventions, large-scale
behavioral changes or population-level depletion of susceptibles have taken
effect~\cite{chowell_bmc2014}.  The exponential growth
rate, $\lambda$, is a function of
epidemiological parameters,
including the transmission rate and ${\cal{R}}_0$~\cite{keeling_2007,wallinga2007generation}.  
For EVD, prior information is available to constrain
the mean duration of the latent phase on the order of 8-12 
days~\cite{eichner2011incubation,teamebola}, the mean infectious
period before death or recovery on the order of 5-9 days~\cite{teamebola,meltzer_2014} and the fraction of disease-induced mortality of 
approximately 70\%~\cite{chowell_2004,teamebola}.   
However, even with these prior constraints, theory does not
predict a one-to-one relationship between ${\cal{R}}_0$ --
the feature we want to infer -- and
$\lambda$ -- the feature that we can measure.
This lack of a
one-to-one relationship gives rise to a so-called identifiability
problem in estimating epidemiological parameters, including
${\cal{R}}_0$, from early-stage epidemic data alone.
Appendix~\ref{app.identify_SIR} presents a rationale for why
identifiability problems arise more generally when fitting
epidemiological models.

To examine the identifiability problem as it pertains to EVD, we 
fit both the SEIR and SEIRD models 
to an exponentially growing epidemic with rate $\hat{\lambda}$
for which the number of cases is increasing with
a characteristic time of $1/\hat{\lambda}=21$ days.
Further,  we assume that 
$T_E=11$ days, 
$T_I=6$ days and $f=0.7$.  We utilize standard epidemiological
methods to infer ${\cal{R}}_0$ for the SEIR model
and, in turn, $\beta_{\mathrm SEIR}$, given observations of $\hat{\lambda}$ (see Eq.~\ref{eq.R0_SEIR}).  We find
a point estimate of $\beta_{\mathrm SEIR}$ of 0.33 and a
corresponding
${\cal{R}}_0$ for the SEIR model of 1.95. Uncertainty in
the duration of periods, risk of mortality
and noise in epidemic case count data 
would lead to corresponding uncertainty
in the value of ${\cal{R}}_0$.  

In contrast, there are three unknown parameters in the SEIRD model:
$\beta_I$, $\beta_D$ and $T_D$.  The time to burial
is, in part, culturally determined, with prior estimates
of 2 days applied to
Ebola outbreaks in Uganda and the Democratic Republic of Congo~\cite{legrand2007understanding} 
having been carried forward to current models (e.g.~\cite{pandey_2014}).
Yet, given the size of the outbreak, additional delays between death
and burial are likely.   Even with a fixed value of $T_D$,
the transmission rates $\beta_I$ and
$\beta_D$ remain unknown.  Hence, trying to fit a SEIRD model to epidemic case
data poses an identifiability problem.
That is to say: there are
potentially many combinations of transmission parameters, $\beta_I$
and $\beta_D$ that can yield the
same observed epidemic growth rate.  
Here, we consider three scenarios, where $T_D=2$, 4 and
6 days.  For each scenario, we must solve for the combination
of $\beta_I$ and $\beta_D$ that yield the epidemic
growth rate $\lambda = 1/21$.  The mathematical details are
in Appendix~\ref{app.R0}.

For a given value of $T_D$, we evaluate a continuum of models 
in which the proportion of ${\cal{R}}_0$ attributable to
post-death transmission varies from 0 to 1.  We define this
fraction as $\rho_D={\cal{R}}_0(\mathrm{dead})/{\cal{R}}_0$.
We find a negative
relationship between the estimated pre- and post-death transmission rate
(compare Figure~\ref{fig.identify_SEIRD}-upper left and middle-left panels).
This negative relationship is a consequence of trying to fit the
same observed case data while modifying the relative
importance of pre- and post-death transmission. Importantly, the point-estimate
of ${\cal{R}}_0$ increases with increasing force of transmission
post-death (Figure~\ref{fig.identify_SEIRD}-lower left).  
Increasing post-death transmission
implies that the average infectious period also increases.
As a consequence, there are fewer
epidemic generations that nonetheless led to the same rise in cases.
This means that
the average number of secondary infections per infected individual
must be higher.  This is a generic feature of epidemiological
models.  The predicted
growth rate for epidemics with these distinct epidemiological
parameters are equivalent -- $\lambda=1/21$
(Figure~\ref{fig.identify_SEIRD}-right) --
despite the differences in underlying rates.

\subsection*{Challenges in fitting early-stage epidemic data of EVD in West Africa due to identifiability problems}
The identifiability problem, described in the previous section,
suggests why it is more difficult than has been recognized
to ascertain the mechanistic details
of EVD transmission from early-stage epidemic data alone.  
Here, we investigate
case data from three countries:
Guinea, Liberia and Sierra Leone (data from~\cite{rivers_github}; see
Table~\ref{tab.case_data} for more information).
We use an exponential growth
curve fit to the cumulative case counts as a target to investigate multiple
possible scenarios (Figure~\ref{fig.identify_SEIRD_data}).
For this fit, we extend our SEIRD model to include a more realistic
distribution period for the E class~\cite{eichner2011incubation,teamebola}.
The exposed (i.e, latently infected)
period is modeled as a gamma distribution with mean of
11 days and 6 classes, so that the standard deviation is
4.5 days (see Supplementary Figure~\ref{fig.exposed_period}).  
We use the generating-function approach of Wallinga and Lipsitch~\cite{wallinga2007generation} (see Appendix~\ref{app.R0_general})
to estimate ${\cal{R}}_0$ from $\lambda$ while accounting for the chosen time distributions within the E, I and D classes.  
For each country, the resulting model
predictions have two key features (see Figure~\ref{fig.identify_SEIRD_data}).  
First, multiple scenarios with varying
ratios of transmission risk from living and dead individuals all fit
the data equally well.  Second, 
estimates that neglect post-death transmission tend to under-estimate ${\cal{R}}_0$. 
The bottom-left
panels of Figure~\ref{fig.identify_SEIRD_data} all show an increase
in ${\cal{R}}_0$ that varies with the fraction of
cases caused by post-death transmission, $\rho_D$.  
The increase in ${\cal{R}}_0$ due to post-death transmission
is of concern.  However, there is a tradeoff: larger $\rho_D$ means not only a larger ${\cal{R}}_0$, but also a larger potential impact of reducing post-death transmission.

\section*{Reduction in transmission risk after death can have substantial epidemiological benefits}
\begin{figure*}[t!]
\begin{center}
\includegraphics[width=0.45\textwidth]{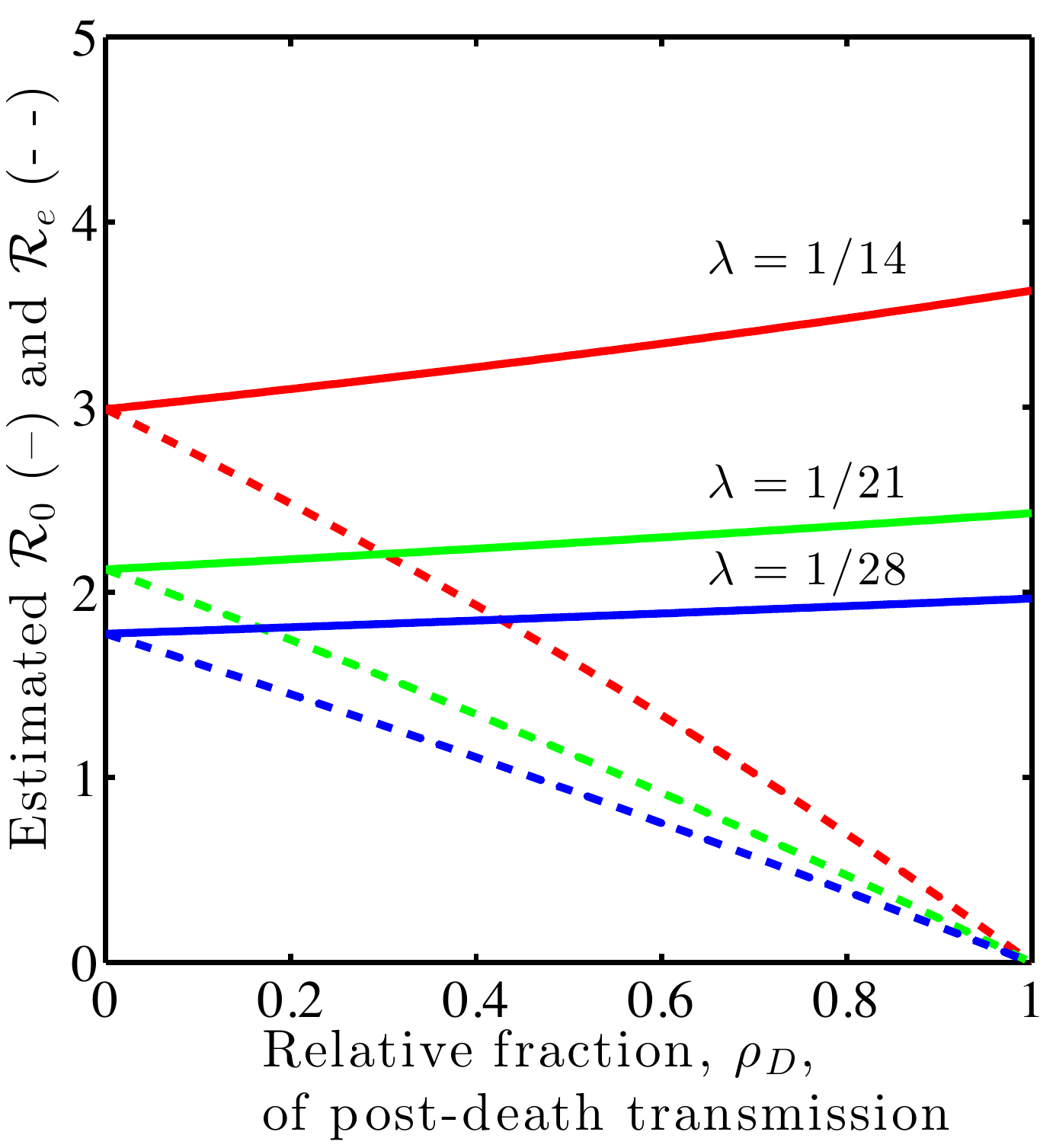}
\mbox{\hspace{0.025\textwidth}}
\includegraphics[width=0.47\textwidth]{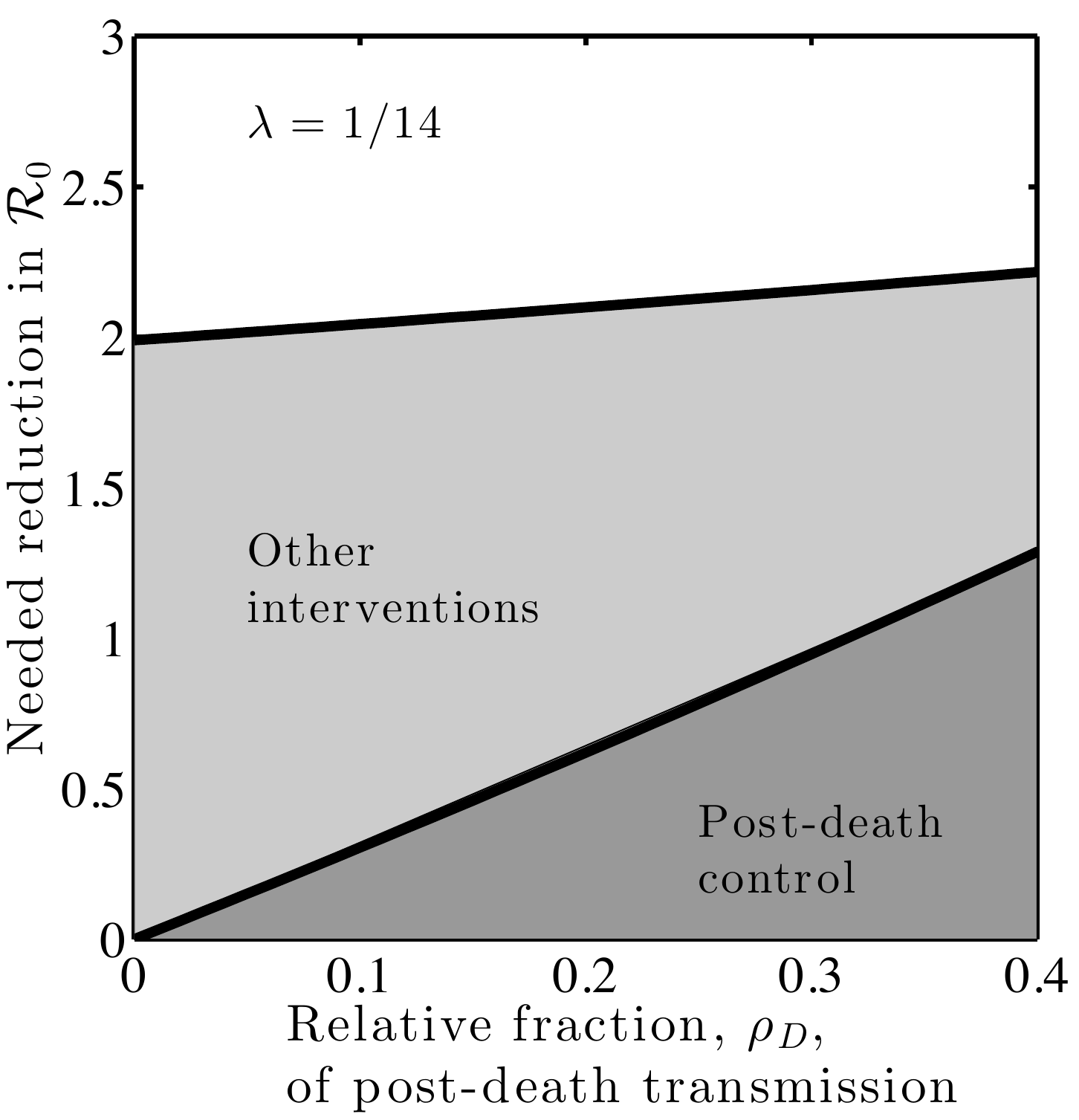}\\
\includegraphics[width=0.47\textwidth]{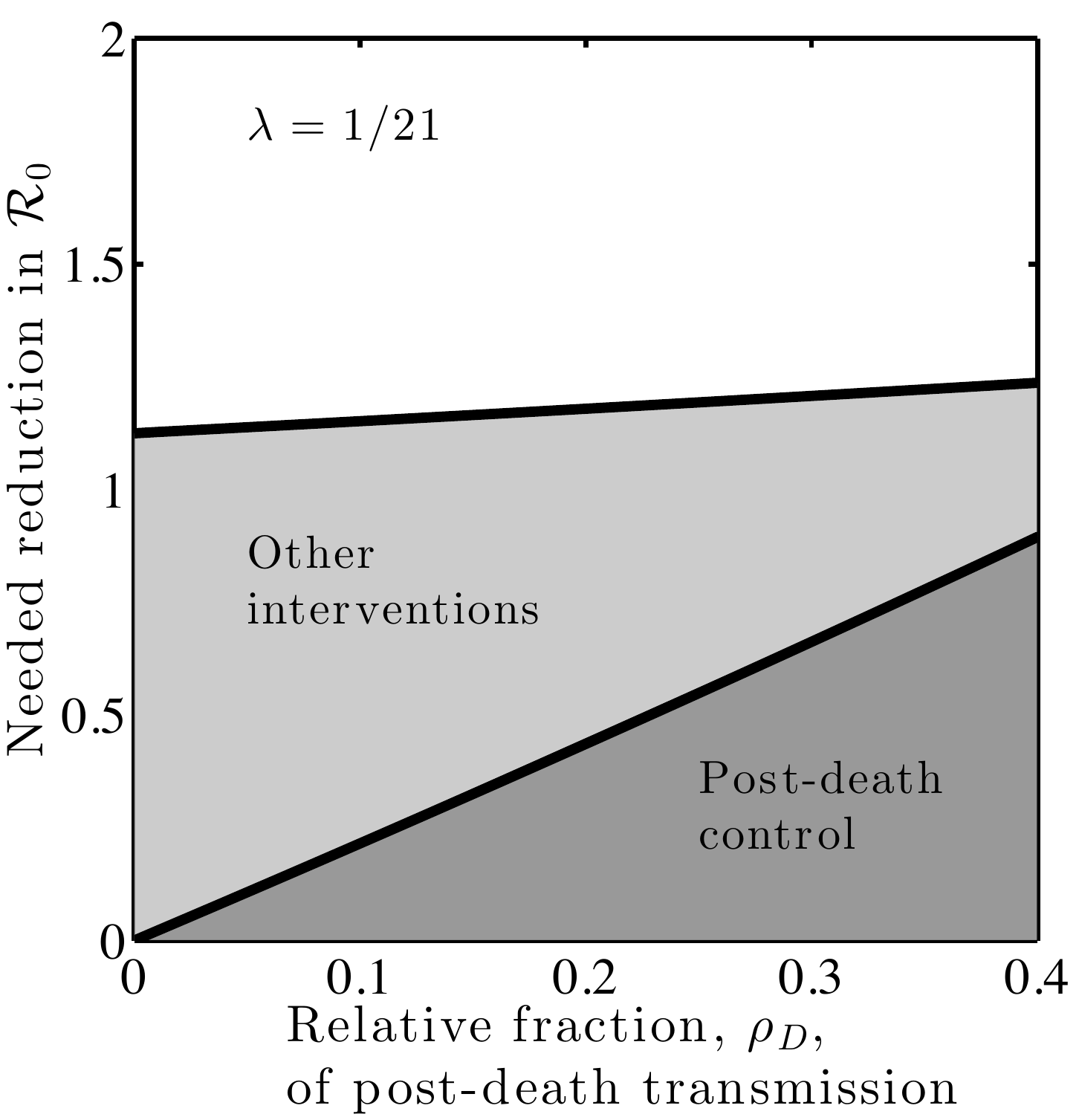}
\mbox{\hspace{0.025\textwidth}}
\includegraphics[width=0.47\textwidth]{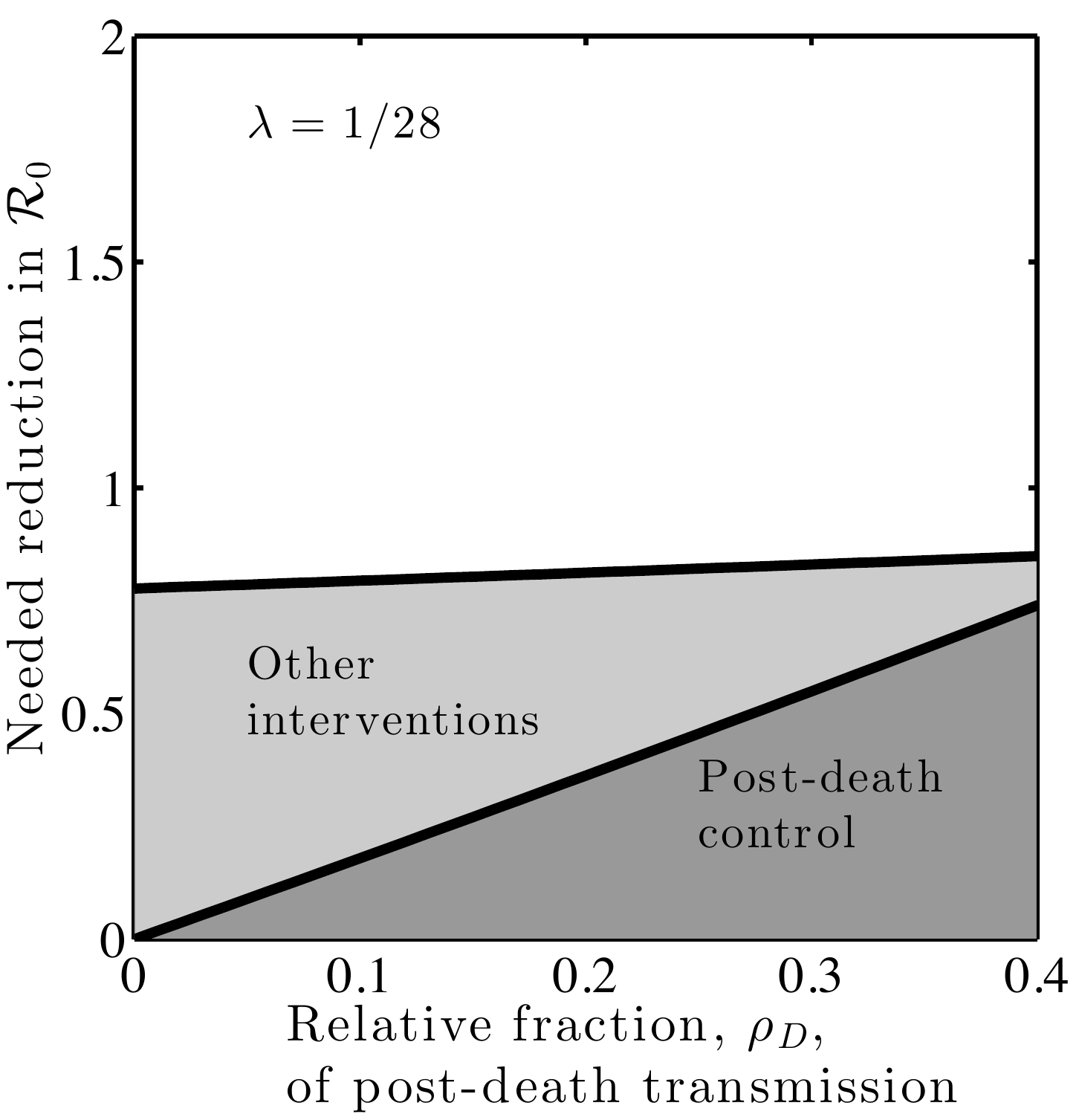}
\caption{
The effect of controlling post-death transmission of EVD outbreaks
with different epidemic growth rates $\lambda$.
(Top-left) Basic reproductive numbers, ${\cal{R}}_0$, without
intervention (solid lines) compared to effective reproductive 
numbers, ${\cal{R}}_e$ by eliminating post-death transmission (dashed lines).
Reproductive numbers are plotted against the fraction
$\rho_D$ of secondary infections due to dead-to-living transmission.
For all scenarios, ${\cal{R}}_e = (1-\rho_D){\cal{R}}_0$.
(Other panels) Break-down of the needed reduction in
${\cal{R}}_0$ to reach a value of ${\cal{R}}_e=1$ for each
of the characteristic epidemic growth rates examined in the top-left panel.  The
dark-shaded region denotes the reduction in secondary
cases due to elimination of post-death transmission
as a function of $\rho_D$.  The light-shaded region
denotes the additional reduction in secondary cases
necessary if post-death transmission is eliminated.
\label{fig.control_R0}}
\end{center}
\end{figure*}

We evaluate the benefits of control  in a
SEIRD representation of EVD using a gamma distributed E class period.
Three scenarios are considered, in which the characteristic
epidemic growth times are $1/\lambda=14$, 21 and 28 days
and for which we assume $T_D=3$ days.  
Figure~\ref{fig.control_R0} summarizes our central findings.
We find, as before, that ${\cal{R}}_0$ is an increasing
function of $\rho_D$, the proportion of transmission that occurs post-death.
We also find that the inferred basic reproductive number increases 
with increasing epidemic growth rates.
These estimates can be used to evaluate
the benefit of control strategies
that eliminate (even partially) post-death transmission.
In the limit that all post-death transmission is eliminated, the effective reproductive number would be ${\cal{R}}_e\equiv (1-\rho_D){\cal{R}}_0$.  
In this limit, ${\cal{R}}_0$ is reduced by $\rho_D{\cal{R}}_0$, a substantial amount  
given estimates of $\rho_D$ in the range of 10\%-30\%~\cite{teamebola}. 
For example, in the scenarios evaluated,
control of post-death transmission reduces ${\cal{R}}_0$ by $\approx 0.2-1$ secondary transmission per infected individual.
Thus, controlling post-death transmission of EVD could be an important component of epidemic control.

\section*{Conclusions}
The relative importance of post-death transmission is difficult to estimate from epidemic growth rate data alone,
and has important implications for estimates of key epidemiological quantities, and for prediction. 
This difficulty is due to what is classically termed an ``identifiability problem'' -
relevant to EVD and to other emerging or poorly characterized infectious diseases.
Despite the challenge in identifying epidemiological parameters, we robustly conclude that neglecting post-death transmission while fitting to epidemic growth rate tends to lead to underestimates of ${\cal{R}}_0$. Such underestimates
are a potential concern for ongoing efforts 
to make realistic models of Ebola dynamics and its control.

Here, we have focused on one feature of such control: the use of burial teams and other practices
intended to reduce post-death transmission.
Burial teams are part of a diverse set of responses required to stop the spread of EVD~\cite{pandey_2014}.  
These responses include behavioral changes, hospital interventions~\cite{meltzer_2014}, and (potentially) vaccination.
More burial teams are needed, but recruiting has proven to be difficult due to stigmatization, lack of personal protective equipment, and insufficient compensation of workers~\cite{reuters_ebola}.
Previous reports have suggested that some individuals in the population experiencing the epidemic may have acquired immunity to Ebola as a result of sub-clinical infections~\cite{leroy2000human,bellan2014ebola}. 
If these individuals exist, and can be identified, they may be valuable contributors to response efforts if they can be recruited as family health-care workers ~\cite{bellan2014ebola} or as part of burial teams.
Moving forward, it is essential to consider the logistics
of deploying burial teams efficiently and safely while
balancing public health benefits and community norms~\cite{hewlett2003,bauch2013}.
A better understanding of post-death transmission can help to understand the EVD epidemic in West Africa and plan control efforts, hopefully leading in the long-term to control and elimination of the current outbreak.

\section*{Acknowledgements}
The authors are indebted to Yao-Hsuan Chen, John Glasser, Brian Gurbaxani, Andrew
Hill and participants in the Modeling Infectious Disease Group at the CDC 
for critical suggestions and feedback that made this effort possible.
The authors thank Hayriye Gulbudak, Luis Jover, Bradford Taylor, and 
Charles Wigington for
critiques of the manuscript. 
JSW is supported by a Career Award at the Scientific Interface from the Burroughs
Wellcome Fund.

\appendix

\clearpage
\newpage
\section{SEIRD model of Ebola dynamics}
\label{app.seird_model}
The SEIRD model includes the dynamics of susceptible,
exposed and infectious individuals, just as in the SEIR model.
It differs in
that the R class stands
for recovered individuals while the D class stands for dead individuals,
who are nonetheless infectious. The dynamics can be written as:
\begin{eqnarray}
\frac{\d S}{\d t} &=& -\beta_I S I/N -\beta_D S D/N  \\
\frac{\d E}{\d t} &=& \beta_I S I/N +\beta_D S D/N -  E/T_E \\ 
\frac{\d I}{\d t} &=& E/T_E - I/T_I \\ 
\frac{\d R}{\d t} &=& (1-f) I/T_I  \\
\frac{\d D}{\d t} &=& f I/T_I - D/T_D \label{eq.D.dynamics}
\end{eqnarray}
where $\beta_I$ is the transmission rate for
contacts with infected individuals and $\beta_D$ is the
transmission rate associated with contacts with dead individuals
(who are nonetheless infectious).  The other parameters are:
$T_E$ the average exposed period,
$T_I$ the average infectious period,
$T_D$ the average period of infectiousness after death
and 
$f$ the fraction of infected individuals who die.
In this model $N=S + E + I + R$.
This model neglects birth/immigration of individuals
and natural death/emigration.
The conventional derivation of the transmission rate 
is that a susceptible individual interacts with
a certain fixed number of individuals per unit time $m$, of which a fraction
$I/N$ are infectious, and only a fraction $p$ of which lead
to transmission.  The transmission rate $\beta_I$ is a product of $m$ and $p$.
Similarly, here we assume that dead individuals are contacted
by a certain fixed number of individuals per unit time $n$,
of which a fraction $S/N$ are susceptible, and only a fraction $q$ of contacts
lead to transmission.  The transmission rate $\beta_D$ is a product
of $n$ and $q$.
The SEIRD model can be extended further to take into account
the possibility that the duration of the exposed
infectious and dead periods are non-exponential.

\section{Estimating the basic reproductive number, ${\cal{R}}_0$, for
the SEIR and SEIRD models given exponential intra-class period distributions}
\label{app.R0}
Infected case data can then be used to estimate unknown epidemiological
parameters, including the transmission rate and ${\cal{R}}_0$.  
In the case of
the SEIR model, the predicted exponential growth rate, $\lambda$,
can be derived from a solution
of the linearized dynamics near the value of $(S=N,E=0,I=0,R=0)$.
The growth rate $\lambda$ correspond to the largest eigenvalue of
the Jacobian:
\begin{equation}
J = \left[
\begin{array}{c c}
-\frac{1}{T_E} & \beta_{SEIR} \\
\frac{1}{T_E} & -\frac{1}{T_I}
\end{array}
\right]
\end{equation}
It can be shown that the growth
of the number of infected cases is an exponential of the
form $I(t)=I_0e^{\lambda t}$ where
\begin{equation}
\lambda = \frac{-(\sigma+\gamma)+\sqrt{\left(\sigma+\gamma\right)^2-4\sigma(\gamma-\beta_{SEIR})}}{2}
\label{eq.seir.lambda}
\end{equation}
where $\sigma \equiv 1/T_E$ and $\gamma\equiv 1/T_I$.
The best-fit value of $\beta_{SEIR}$ can be inferred 
given a measured value $\hat{\lambda}$ and prior estimates for
$\sigma$ and $\gamma$.  For example, 
\begin{equation}
{\cal{R}}_{0,SEIR} = (1+\lambda/\sigma)(1+\lambda/\gamma)
\label{eq.R0_SEIR}
\end{equation}
where ${\cal{R}}_{0,SEIR} = \beta_{SEIR}/\gamma$, such that
\begin{equation}
\hat{\beta}_{SEIR} = (1+\hat{\lambda}T_E)(1+\hat{\lambda}T_I)/T_I
\end{equation}

Similarly, given the SEIRD model, 
the predicted exponential growth rate $\lambda$ of the number
of infected cases corresponds to the largest eigenvalue of
the linearized system near $(N,0,0,0,0)$, of which only the variables
$E(t)$, $I(t)$ and $D(t)$ must be tracked.  The Jacobian of this
subsystem is:
\begin{equation}
J = \left[
\begin{array}{c c c}
-\frac{1}{T_E} & \beta_I & \beta_D \\
\frac{1}{T_E} & -\frac{1}{T_I} & 0 \\
0 & \frac{f}{T_I} & -\frac{1}{T_D}
\end{array}
\right]
\end{equation}
The solutions can be computed exactly.

\section{The roots of identifiability problems in estimating the
basic reproductive number from early-stage epidemic growth data}
\label{app.identify_SIR}
\begin{figure*}[t!]
\begin{center}
\includegraphics[width=0.95\textwidth]{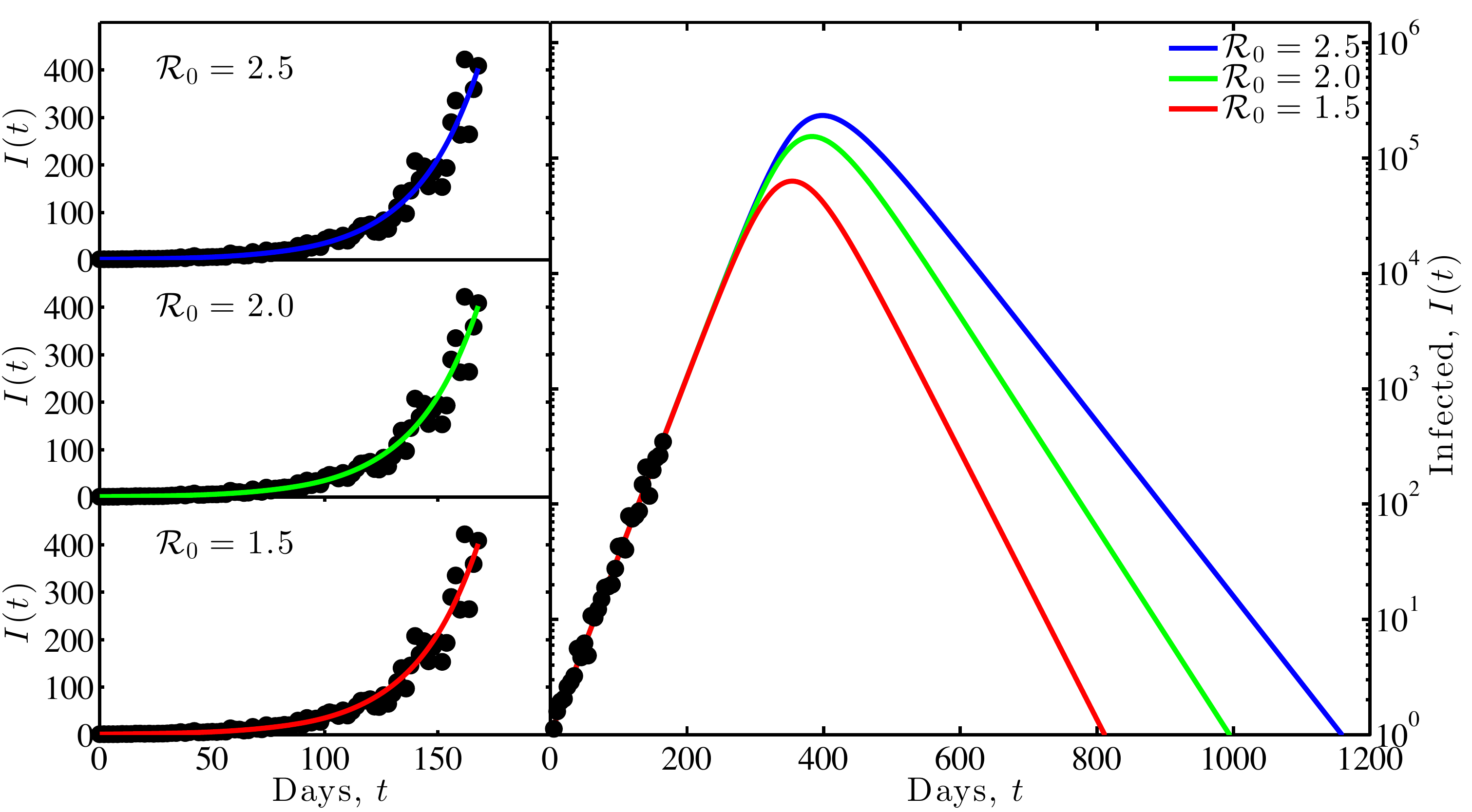}
\caption{Identifiability problem in estimating ${\cal{R}}_0$ for a SIR model from
exponential epidemic growth data.  The synthetic data (black circles)
is $I(t)\propto e^{\lambda t}e^{1+\psi}$ where $\lambda=1/28$, corresponding
to a characteristic time of 4 weeks and 
where $\psi$ is a normally distributed random variable with mean 0
and standard deviation 0.1.  The model fits
correspond to solutions of SIR models in which
$\beta=0.107,$ $0.0714$, and $0.0595$ days$^{-1}$
and
$T_I=14$, 28 and 42 days respectively.  The basic reproductive
number in each case is ${\cal{R}}_0 = \beta T_I$ = 1.5,
2.0 and 2.5 respectively.  (Left panel) Each of the SIR
model predictions fits the data equally well at early times, 
despite having very different basic reproductive numbers.
(Right panel) The predictions of the long-term dynamics
differ, with epidemic size increasing as a function of ${\cal{R}}_0$.
\label{fig.identify_SIR}}
\end{center}
\end{figure*}
The identifiability problem raised in the main text is a generic
issue in epidemiology.  By means of illustration, 
consider the spread of a disease that has no exposed stage, such that
it can be suitably described using a SIR model. Further assume
that the basic reproductive number of the disease is to be
estimated from epidemic case data in which the number of cases is 
growing at a rate of $\hat{\lambda}=1/28$.
The basic reproductive number for a SIR
model is $\beta T_I$, i.e., the transmission rate multiplied
by the infectious period.  The epidemic growth rate for a SIR model
is $\lambda = \beta - 1/T_I$, i.e., the difference between
the transmission and recovery rate.  This can be written as:
$\lambda = T_I({\cal{R}}_0-1)$.  Hence, consider three scenarios,
in which the true infectious period is $T_I = 14$, 28 and 42 days.
Each of these scenarios is compatible with the same epidemic growth
rate $\hat{\lambda}$ given ${\cal{R}}_0 = 1.5$, 2.0 and 2.5.  
Figure~\ref{fig.identify_SIR} illustrates this point using synthetic data.
Note that for a given epidemic growth rate, diseases whose period
of infectious is longer have larger basic reproductive numbers.  
In the example above, a disease with
an infectious period of 14 days requires 2 infection cycles (on average)
to increase in case count by a factor of e (2.718).  
Whereas, a disease with an infectious period
of 28 days requires 1 infection cycle (on average) to increase in case count
by a factor of e (2.718).
This is the intuition behind the seemingly paradoxical result that
diseases with longer infectious periods are estimated to have higher values of ${\cal{R}}_0$
when estimated via the same epidemic growth rate.
Moreover, although the disease dynamics may appear indistinguishable at early stages
of an epidemic, the long-term dynamics can be quite different. For
example, Figure~\ref{fig.identify_SIR} shows how 
diseases with higher values of ${\cal{R}}_0$
infect more people over the long-term despite having the same
early time dynamics.  Controlling a disease with
a higher value of ${\cal{R}}_0$ is also more difficult.

\section{Estimating the basic reproductive number, ${\cal{R}}_0$, for
the SEIRD model given arbitrary intra-class period distributions}
\label{app.R0_general}
Wallinga and Lipsitch~\cite{wallinga2007generation} 
established
a formal connection between ${\cal{R}}_0$ and the epidemic 
growth rate, here: $\lambda$, such that
\begin{equation}
{\cal{R}}_0 = \frac{1}{M(-\lambda)}
\end{equation}
where 
\begin{equation}
M(z)=\int_0^\infty e^{za}g(a)\d a
\end{equation}
The moment generating function $M(z)$ operates on
the distribution $g(a)$ which, in epidemiological terms,
is the normalized fraction of all secondary cases caused
by an infectious individual at ``age'' $a$ since infection.
For example, if individuals are only infectious
at a single age $a_c$ after infection, then $g(a)=\delta(a-a_c)$ where
$\delta(x)$ is the delta function.  Similarly, if individuals
recover from being infected at a rate $\gamma$, then
$g(a)=\gamma e^{-\gamma a}$, i.e., an exponential distribution.
The advantage of this approach is that it is possible
to uniquely identify the value of ${\cal{R}}_0$
given a measured epidemic growth
rate $\hat{\lambda}$ and additional information on the 
age distributions for secondary infections.

\begin{figure}
\begin{center}
\includegraphics[width=0.95\columnwidth]{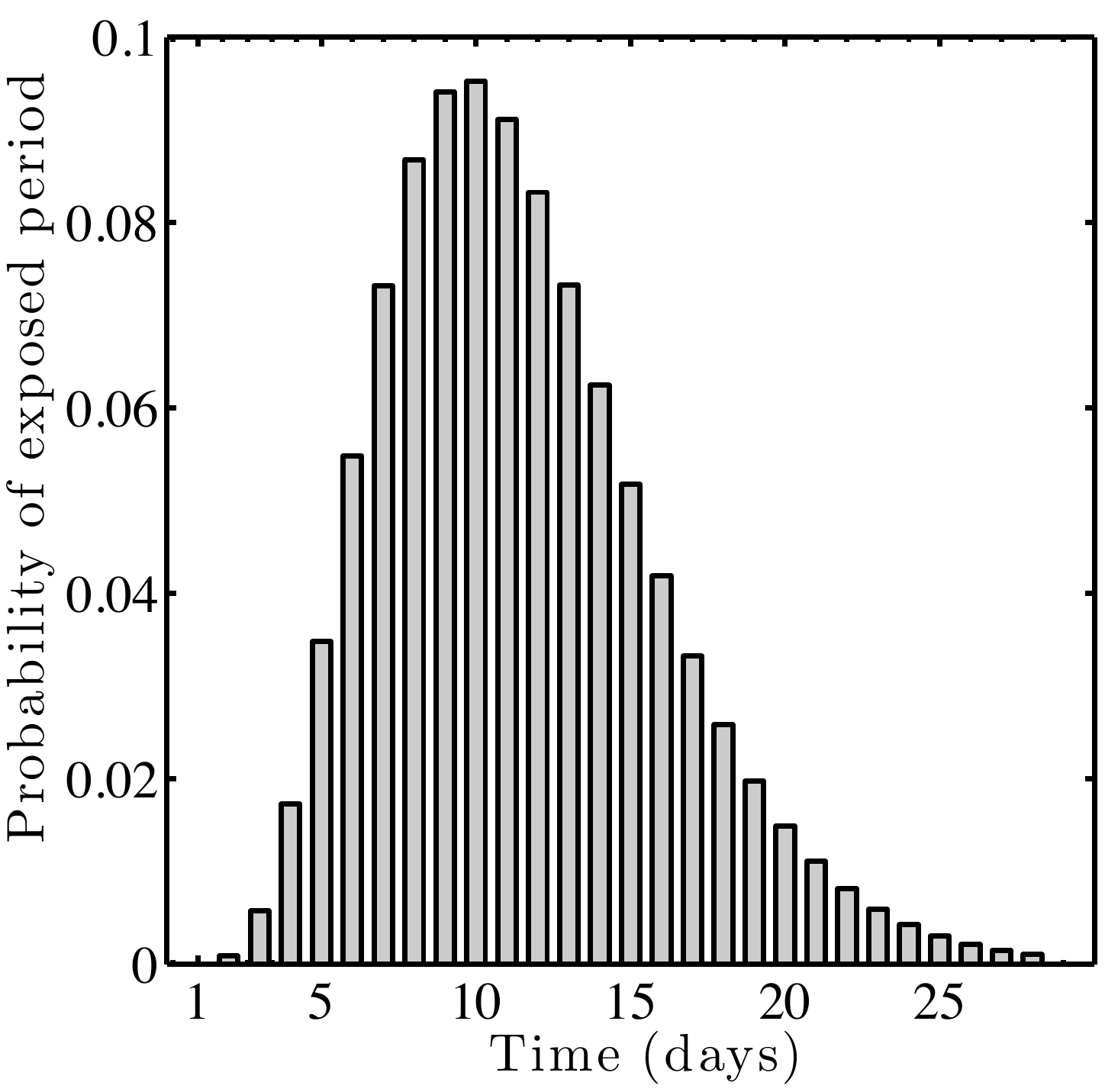}
\caption{Gamma-distributed exposed period.
The distribution has an average exposed period with $T_E=11$ days,
such that the shape parameters are $n_E=6$ and $b_E=n_E/T_E$.
\label{fig.exposed_period}}
\end{center}
\end{figure}

For the SEIRD model, the appropriate generating function is:
\begin{equation}
M(z) = (1-\rho_D)M_E(z) M_I(z)+\rho_D M_E(z)M_I(z)M_D(z)
\end{equation}
where $\rho_D$ is the fraction of secondary transmission due
to post-death transmission and $1-\rho_D$ is the fraction of secondary
transmission due to pre-death transmission.  We consider
a gamma distributed exposed period with $T_E=11$ days,
and shape parameters $n_E=6$ and $b_E=n_E/T_E$ (see Figure~\ref{fig.exposed_period})
whose generating function is:
\begin{equation}
M_E(-\lambda) = \left(\frac{b_E}{b_E+\lambda}\right)^{n_E}
\end{equation}
We consider here exponentially-distributed periods for the I and D classes.
The generating functions are:
\begin{eqnarray}
M_I(-\lambda) &=& \frac{\gamma}{\gamma+\lambda}\\
M_D(-\lambda) &=& \frac{\chi}{\chi+\lambda}
\end{eqnarray}
where $\gamma=1/T_I$ and $\chi=1/T_D$.
Therefore for the SEIRD model, it is possible to estimate
${\cal{R}}_0$ using the generating function method
given observations of an epidemic growth rate
and suitable information on epidemiological modes and parameters.

This analysis assumed
that the I and D classes are exponentially distributed with
characteristic times of 6 and 3 days, respectively.  A similar
analysis can be performed
in which the force of transmission is concentrated with
different distributions, e.g., uniform, unimodal or even
concentrated
at the very end of a fixed epidemic period (so-called delta
distributed).  

\section{Case data information}
\label{app.sec.data}
Cumulative case count data from Guinea, Liberia
and Sierra Leone was used as the target of model fits in Figure~\ref{fig.identify_SEIRD_data}.
These data sets were downloaded from Caitlin Rivers' publicly available
github site~\cite{rivers_github}.  The time periods
over which data is calibrated is shown in Table~\ref{tab.case_data}.
The start date was selected based on the first day at which the
cumulative case count exceeded 50.  The final date was set at the end
of August, coinciding with reported increases in intervention and
widespread dissemination of the severity of the outbreak~\cite{pandey_2014}.
The exponential fits are based on log-transformed cumulative case counts.
Additional challenges for inference arise, in part,
due to under-reporting~\cite{meltzer_2014} and lags between incidence
and reporting events~\cite{teamebola}. 

\begin{table}[h!]
\begin{center}
\begin{tabular}{l|llll}
Country & $T_0$ & $T_1$ & $\hat{\lambda}$ & Doubling period \\
\hline\hline
Guinea  & 3/22/14 & 8/31/14 & 0.011 & 61 days \\
Liberia & 6/22/14 & 8/31/14 & 0.048 & 14 days \\
Sierra Leone & 5/28/14 & 8/31/14 & 0.032 & 21 days
\end{tabular}
\caption{Data sources for model fits of SEIRD to Ebola epidemic
data.  The values of $T_0$ and $T_1$ denote the start
and stop dates for the cumulative case data used for estimating
the epidemic growth rate, $\hat{\lambda}$. Estimates of
the epidemic growth rate were based on linear regression of
log-transformed cumulative case counts.  The doubling
time of the epidemic is defined as $\frac{\log{2}}{\hat{\lambda}}$.
\label{tab.case_data}}
\end{center}
\end{table}

\end{document}